\documentclass{ws-procs9x6}            

	\usepackage{braket}
	\usepackage[utf8]{inputenc}

\begin{document}
\title{Partial Wave Mixing in Hamiltonian Effective Field Theory}

\author{Yan Li$^{*,a}$, Jia-jun Wu$^a$, Curtis D. Abell$^b$, Derek B. Leinweber$^b$\\ and Anthony W. Thomas$^{b,c}$}

\address{$^a$School of Physical Sciences, University of Chinese Academy of Sciences (UCAS), Beijing 100049, China\\
$^b$Special Research Center for the Subatomic Structure of Matter (CSSM),
Department of Physics, University of Adelaide, Adelaide 5005, Australia\\
$^c$ARC Centre of Excellence for Particle Physics at the Terascale (CoEPP),
Department of Physics, University of Adelaide, Adelaide, South Australia 5005, Australia\\
$^*$E-mail: liyan175@mails.ucas.edu.cn\\
}

\begin{abstract}
	We explore partial-wave mixing in the finite volume based on HEFT, and provide the P-Matrix to show the degree of partial-wave mixing. An example of isospin-2 $\pi\pi$ scattering is used to check the consistency between HEFT and L\"{u}scher's method.
\end{abstract}


\bodymatter

\section{Introduction}
	The infinite-volume phase shift observed in experiment and the finite-volume spectrum obtained from Lattice QCD proved to be related by the model independent L\"{u}scher's formula \cite{Luscher:1985dn,Luscher:1986pf,Luscher:1990ux}.
	Another equivalent approach connecting them is Hamiltonian effective field theory (HEFT) \cite{Hall:2013qba,Hall:2014uca,Liu:2015ktc,Wu:2017qve}.
	In HEFT, one has the infinite- and finite-volume Hamiltonians parameterized by a common set of parameters. The phase shift can be predicted through the Hamiltonian constrained by the Lattice QCD spectrum.

	As a consequence of the violation of spherical symmetry in the finite volume, the partial wave quantum numbers $(l,m)$ fail to be good ones.
	That different partial waves are mixed in the finite volume is called partial-wave mixing. Here we explore partial-wave mixing in HEFT, and provide the P-Matrix defined in \eref{eq:PM} to reflect the degree of partial-wave mixing. Finally, an example of isospin-2 $\pi\pi$ scattering is used to examine the consistency between this method and L\"{u}scher's method.

\section{Partial Wave Mixing in HEFT}
	In the infinite volume, to utilize the spherical symmetry, one combines the plane wave states $\ket{\mathbf{k}}$ with spherical harmonics $Y_{l,m}(\hat{\mathbf{p}})$ to define
	\begin{equation}\label{eq:klm}
		\ket{k;l,m}:=\int d\Omega_{\hat{\mathbf{k}}}\,Y_{lm}(\hat{\mathbf{k}})\,\ket{\mathbf{k}} \,.
	\end{equation}
	Then spherical symmetry only allows interactions of the following form
	\begin{equation}
		\hat{V} = \int\frac{k'^2\,dk'}{(2\pi)^3}\int\frac{k^2\,dk}{(2\pi)^3}\,\sum_{l,m}\,v_l(k',k)\,\ket{k';l,m}\bra{k;l,m} \,.
	\end{equation}

\begin{figure*}[h]
		\centering
		\includegraphics[width=10cm]{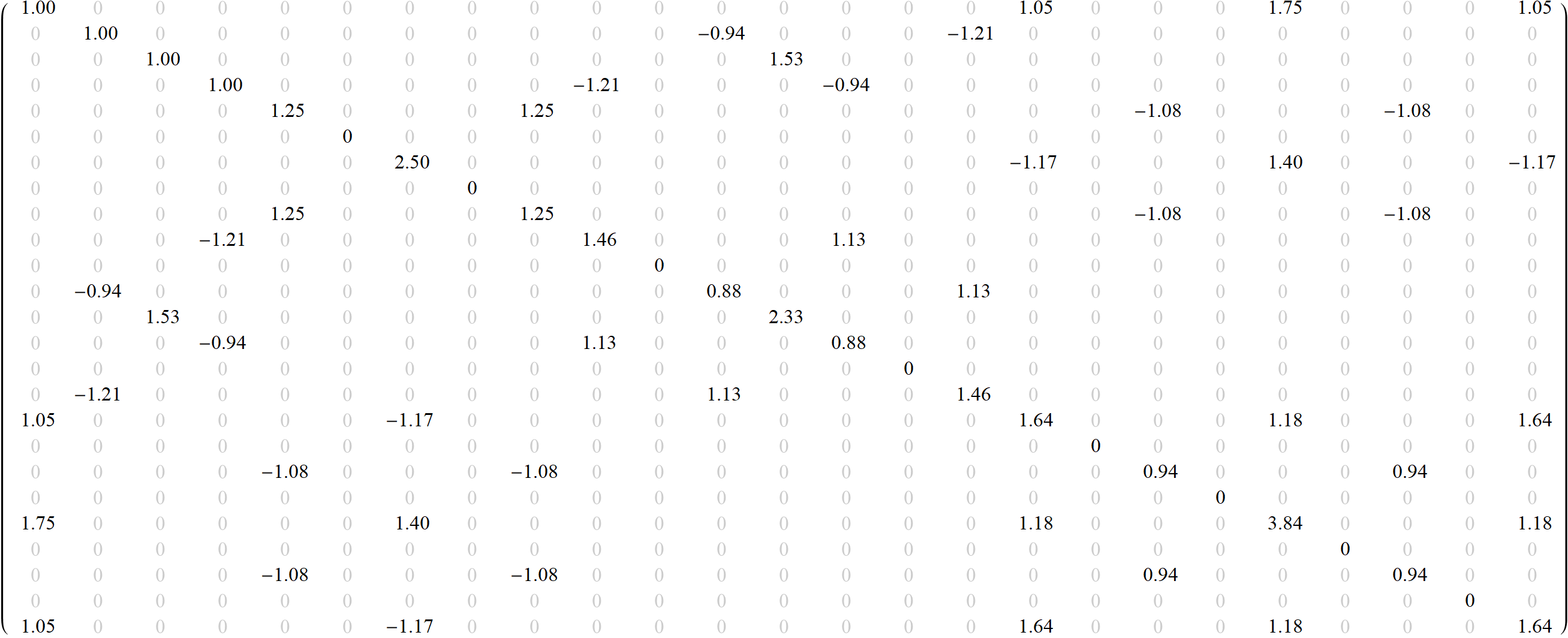}

		\vspace{.5cm}

		\includegraphics[width=10cm]{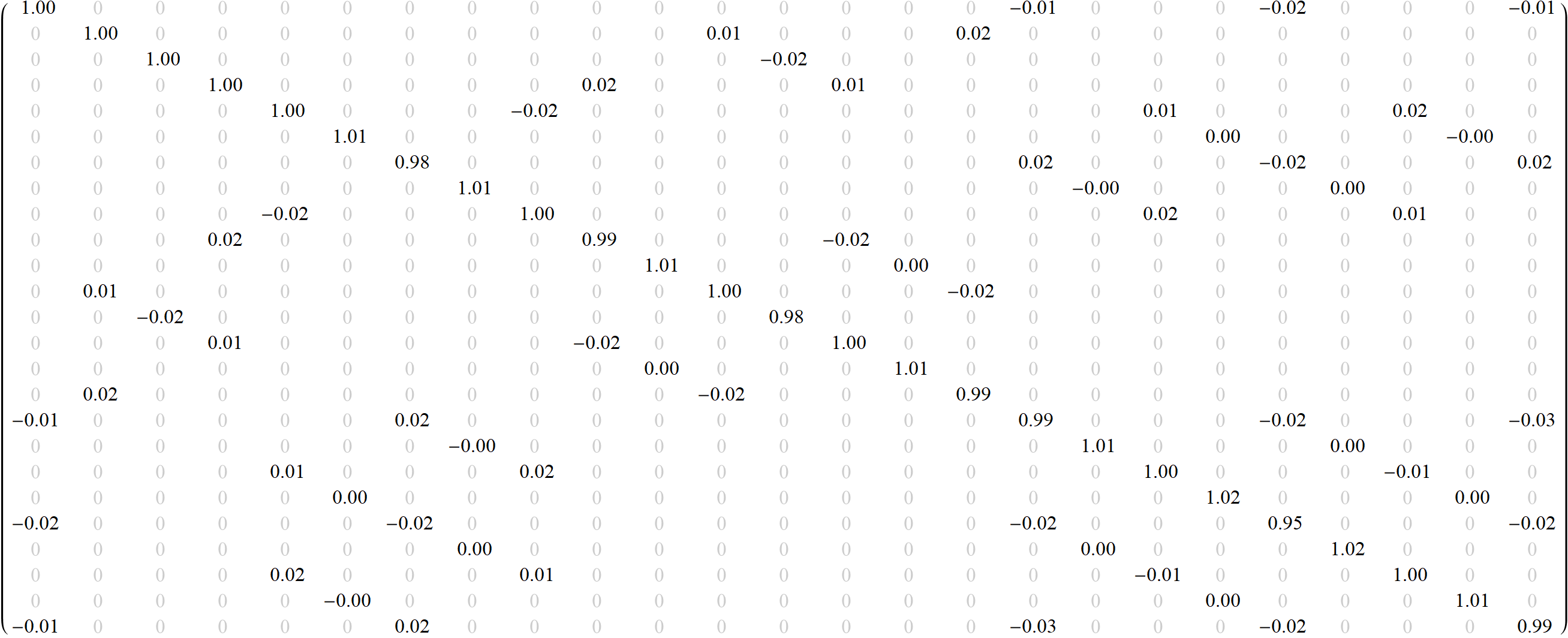}
		\caption{$P_{N}/C_3(N)$ with $N=1,\,941$ and $C_3(N)=6,\,552$ respectively: The $25\times25$ matrix ordered as $(l,m) =
		  (0,0),\,(1,-1),\,(1,0),\,(1,1),\,\cdots,\,(4,4).$}
		\label{fig:PM}
	\end{figure*}

	In the finite volume, only the periodic plane wave states denoted by $\ket{\mathbf{n}}$ will survive. Therefore the integration in \eref{eq:klm} should be replaced by a summation of discrete momentum. Furthermore, one can define new states in the finite volume as follows
	\begin{equation}
		\ket{N;l,m} = \sum_{|\mathbf{n}|^2=N}\sqrt{4\pi}\, Y_{lm}(\hat{\mathbf{n}})\ket{\mathbf{n}} \,.
	\end{equation}
	That $\ket{N;l,m}$ are in general not orthogonal to each other reflects the partial-wave mixing. One can define the inner product matrix (named as P-Matrix)
	\begin{equation}\label{eq:PM}
		[P_N]_{l',m';l,m} := \braket{N;l',m'|N;l,m} = 4\pi \sum_{|\mathbf{n}|^2=N} Y_{l'm'}^*(\hat{\mathbf{n}})Y_{lm}(\hat{\mathbf{n}})
	\end{equation}
	to show the degree of partial-wave mixing.
	If one considers a finite but non-zero momentum $2\pi\sqrt{N}/L$, in the infinite-volume limit $L\to\infty$, one can see the recovery of the spherical symmetry as follows
	\begin{equation}
		[P_N]_{l',m';l,m} \xrightarrow{N\to\infty} C_3(N)\,\delta_{l',l}\,\delta_{m',m} \,,
	\end{equation}
	where $C_3(N)$ counts the $\mathbf{n}$ satisfying $|\mathbf{n}|^2=N$.
	Here we provide a few examples of the numerical values for $P_N/C_3(N)$ with $N = 1$ and $941$ in \fref{fig:PM}, and it is indeed approaching to an identity matrix.

	To make use of the finite-volume symmetry, one combines $\ket{N;l,m}$ into states that respect the cubic symmetry
	\begin{equation}
		\ket{N,l;\Gamma,f,\alpha} = \sum_m\,[C_l]_{\Gamma,f,\alpha;m}\ket{N;l,m} \,,
	\end{equation}
	where $(\Gamma,f,\alpha)$ represents the $\alpha$-th vector of the $f$-th occurrence (in a given $l$) of the irreducible representation $\Gamma$, and $[C_l]_{\Gamma,f,\alpha;m}$ are group theoretic constants independent of $N$.
	The inner products between $\ket{N,l;\Gamma,f,\alpha}$ relate to the P-Matrix directly from the definition. The states after orthonormalization will be the final basis.

	Working with $\ket{\mathbf{n}}$, one uses a momentum cutoff $N_{\text{cut}}$ to make the finite-volume Hamiltonian of finite dimension. With $\ket{N;l,m}$, one then uses a partial wave cut $l_{\text{cut}}$ to decouple the potential from high partial waves. With $\ket{N,l;\Gamma,f,\alpha}$, different irreducible representations of the cubic group are also decoupled. Finally, the final dimension of the finite-volume Hamiltonian will be significantly reduced. 

\section{Example of Isospin-2 {\large $\mathbf{\pi\pi}$} Scattering}\label{sec:EIS}
	\begin{figure*}[h]
		\centering
		\includegraphics[width=5cm]{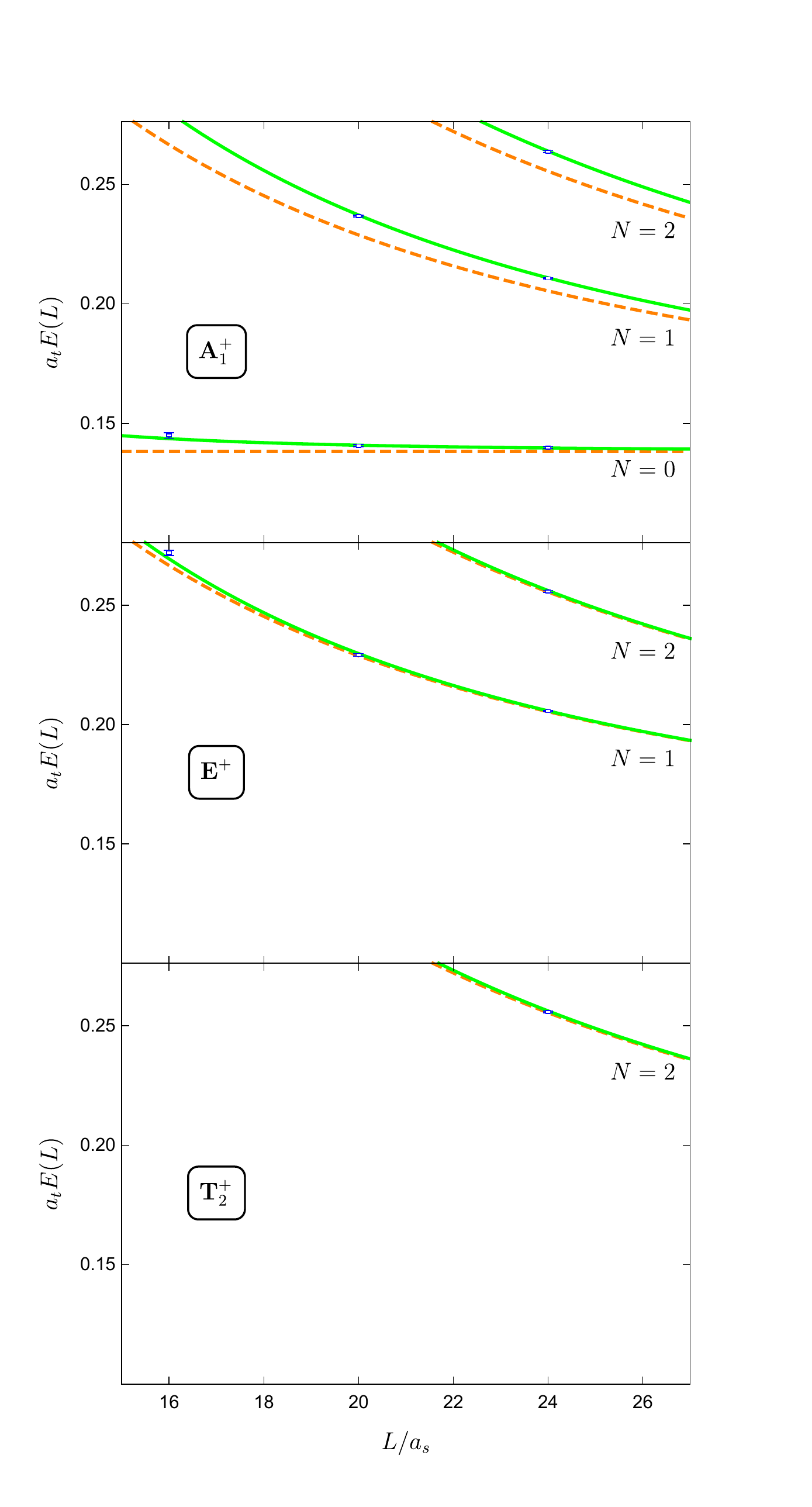}
		\includegraphics[width=5cm]{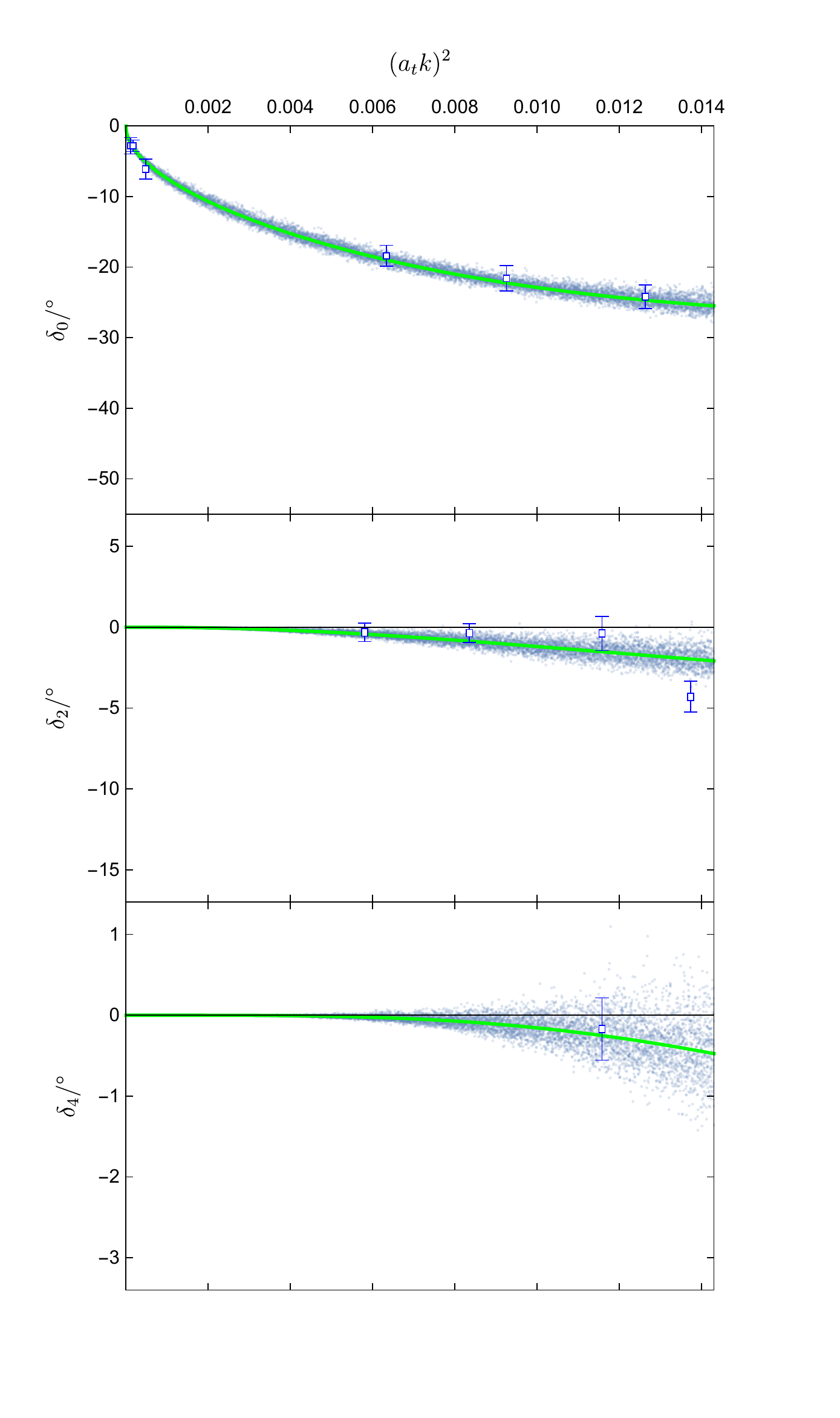}
		\caption{Left: Data points describe the lattice spectrum for irreps $\mathbf{A}_1^+$, $\mathbf{E}^+$, $\mathbf{T}_2^+$ -- from Ref.~\citenum{Dudek:2012gj}. Dashed curves represent the non-interacting rest-frame pion-pair energies $2\sqrt{m_\pi^2+k_N^2}$. Solid curves represent the HEFT prediction of the volume dependent spectrum using the fitted parameters. Right: Data points are phase shifts predicted by L\"{u}scher's method -- from Ref.~\citenum{Dudek:2012gj}. Solid curves represent the HEFT prediction of the $s$- (top), $d$- (middle) and $g$- (bottom) wave phase shifts and the scattered points describe the uncertainty.}
		\label{fig:Ex}
	\end{figure*}
	
	The parametrization for the potential is chosen to be (measured in units of lattice spacings)
	\begin{equation}
		v_l(p,k)=f_l(p)\,G_l\,f_l(k) \,,\qquad f_l(k) = \frac{(d_l\,k)^{l}}{\left(1+(d_l\,k)^2\right)^{l/2+2}} \,,
	\end{equation}
	where there are 2 parameters $G_l$ and $d_l$ for each partial wave.
	The partial wave cut $l_{\text{cut}}=4$ and Bose symmetry only allow $s$-, $d$- and $g$-waves. The lattice QCD results from Ref.~\citenum{Dudek:2012gj} and the HEFT results are shown in \fref{fig:Ex}, in which HEFT and L\"{u}scher's method are indeed consistent.

\section{Summary}
	We have explored partial-wave mixing in the finite volume based on HEFT, and provided the P-Matrix defined in \eref{eq:PM} to show the degree of partial-wave mixing. Then an example of isospin-2 $\pi\pi$ scattering was used to check the consistency between HEFT and L\"{u}scher's method.

\section*{ACKNOWLEDGEMENTS}
	It is a pleasure to thank Stephen Sharpe for interesting discussions on the research presented
	herein during his visit as a George Southgate Fellow.  The finite-volume energy levels and their
	covariances from Ref.~\citenum{Dudek:2012gj} were provided by the Hadron Spectrum Collaboration -- no
	endorsement on their part of the analysis presented in the current paper should be assumed.
	This project is also supported by the Thousand Talents Plan for Young Professionals.
	This
	research was supported by the Australian Research Council through ARC Discovery Project Grants
	Nos.\ DP150103101 and DP180100497 (A.W.T.) and DP150103164 and DP190102215 (D.B.L.).

\bibliographystyle{ws-procs9x6} 
\bibliography{references}

\end{document}